\documentclass[12pt]{article}
\pdfoutput=1
\usepackage{mathrsfs}
\usepackage[table]{xcolor}
\usepackage{cite}
\usepackage{avant}
\usepackage{amssymb}
\usepackage{multirow,booktabs}
\usepackage{color}
\usepackage{psfrag}
\usepackage{graphicx}
\usepackage{subfigure}
\usepackage{rotating}
\usepackage{cite}
\usepackage[colorlinks=true,linktocpage=true,linkcolor=blue,citecolor=blue]{hyperref}
\usepackage[section]{placeins}
\usepackage{amsfonts}
\usepackage{ifpdf}
\usepackage[toc,page]{appendix}
\usepackage{caption}
\usepackage{pifont}
\usepackage{amsmath}
\usepackage[mathscr]{eucal}
\usepackage[all]{xy}
\usepackage{listing}
\usepackage{pdfpages}
\usepackage{graphicx}
\xyoption{arc}
\xyoption{curve}
\numberwithin{equation}{section}
		\newfont{\spfnt}{punk12}
		\newcommand\email[4]{#1@#2.#3.#4}

		\newcommand\be{\begin{equation}}
		\newcommand\ben{\begin{equation}\nonumber}
		\newcommand\ee{\end{equation}}
		
		\def\pa{\partial}

		\newfont\sheafnt{rsfs10}

\begin{document}
			\title{Charged AdS black holes with higher derivative corrections in presence of
string cloud}
			\author{
			Tanay K. Dey\thanks{\email{tanay.dey}{gmail}{com}{} and \email{tanay.d}{smit}{smu}{edu}{in}}
				~and
				Subir Mukhopadhyay\thanks{\email{subirkm}{gmail}{com}{}}
			}
			\date{
			$^*$ {\small Department of Physics, Sikkim Manipal Institute of Technology, Sikkim Manipal University,
Majitar, Rongpo, East Sikkim, Sikkim-737136, India.}\\
			$^\dagger$ {\small Department of Physics, Sikkim University, 6th Mile, Gangtok 737102..}
			 }
			\maketitle
			\begin{abstract}
				\small
				\noindent
				We have obtained charged asymptotically AdS black hole solutions in Einstein Gauss Bonnet gravity in presence of string cloud. It turns out that there exist three black holes in certain range of parameters for small enough chemical potential. However, for large chemical potential, these black holes merge into a single one. The free energy of the black holes is computed from the on shell Euclidean action by subtracting the contribution of an extremal black hole. We have studied the free energy and specific heat of the solutions, which shows that the medium sized black hole is unstable. 
As an application, we have studied the quark-antiquark potential in the dual gauge theory. This study provides necessary ingredients for study of the dual theories in the context of condensed matter systems.

			\end{abstract}
			\thispagestyle{empty}
			\clearpage
\section{Introduction}

AdS/CFT duality has emerged as an extremely successful approach to study strongly coupled theories. In its basic formulation, it relates ${\mathcal N}=4$ super Yang-Mills theory (SYM) to gravity theory on anti de Sitter (AdS) space living in one higher dimensional space time\cite{ Maldacena, Witten1, Witten2}. In particular, there is a correspondence between the black hole configuration and the deconfined phase of the gauge theory while the pure AdS corresponds to the confined phase. Pure AdS to black hole transition can be interpreted as confinement-deconfinement transition\cite{H-P}. 

Fundamental quarks can be incorporated in the gauge theory by considering cloud of hanging strings in the gravity theory \cite{Guen, Karch, Head, Big, Kumar}. The endpoints of the hanging string traces out contour of the loop at the asymptotic boundary and the endpoints represent quark and antiquark in the dual field theory. A number of works have appeared where gravitational configurations with string cloud have been studied \cite{Chakrabortty:2011sp, tanay1, Hers, Ghaff, Ghosh, Ghosh:2014pga,Ghan1, Ghan2}. Einstein gravity in presence of string cloud admits three black hole solutions and its thermodynamics has been studied and a Hawking-Page like transition has been found between two black holes \cite{tanay1}.

So far the UV completion of the Einstein theory  coupled with matter is concerned, higher derivative corrections are the common features, which get suppressed where the scale of the new effective description becomes relevant.
Out of possible such corrections, Gauss-Bonnet (GB) term turns out to be quite interesting. In four dimension it is a topological term. In other dimensions it has been found that no ghost-like propagating modes appeared with modified kinetic terms\cite{boulware,zwiebach}
at the level of Ricci squared corrections. The Gauss-Bonnet term also appears as the $\alpha^\prime R^2$ correction in the Heterotic string theory \cite{gross,berg,chem} and it may appear in the compactified string theories as well.
From the perspective of duality, these corrections leads to $\frac{1}{\sqrt{\lambda}}$ corrections in the `t Hooft coupling in the gauge theory. Therefore, from the perspective of AdS/CFT duality, solutions in the higher derivative gravity theory can shed light on the subleading corrections in the `t Hooft coupling.

Thermodynamics of Einstein-Gauss-Bonnet gravity has been discussed in a number of works\cite{Cvetic:2001bk, Nojiri:2001aj, Nojiri:2002qn, Cai, tanay2} in absence of string cloud. In presence of string cloud, thermodynamics of black holes in Einstein-Gauss-Bonnet gravity has been explored in \cite{Hers,Ghaff,Dey:2020yzl}. Also, solution and thermodynamic study of black holes (in presence of string cloud) in third order Lovelock gravity has  been discussed in \cite{Ghosh:2014pga}.
 However, in these theories  in presence of string clouds electromagnetically charged solution has not been considered. From the viewpoint of AdS/CFT correspondence, the gravitationl dual of finite density configurations correspond to charged black hole. Therefore, such a solution will enable us to study subleading corrections (in `t Hooft coupling) in the dual theory (with the effect of string cloud). 

In the present work we have obtained a charged black hole sp;ution in Einstein Gauss Bonnet gravity coupled to the Maxwell action in presence of a string cloud. 
In presence of string cloud we find existence of three black holes in certain range of parameter for small enough chemical potential. It turns out that as the chemical potential increases these black holes merge into a single black hole. 

Thermodynamics of charged black hole in Einstein Gauss Bonnet gravity has been studied in \cite{Anninos:2008sj,Dey:2007vt} in absence of string cloud.We have computed the free energy from the Euclidean action. In presence  
of string cloud, in order to regularise the free energy we subtract the contribution of an extremal black hole. In order to analyse the stability of different black holes we have studied its free energy and specific heat. As an application we have studied the quark-antiquark potential in the dual gauge theory by using standard machinery of AdS/CFT. This study furnish an arena for the study of the dual theories in the context of condensed matter systems, quark-gluon plasma etc..

The paper is structured as follows. In the next section we have discussed the black hole solution in Einstein Gauss Bonnet gravity in presence of string cloud. Section 3 deals with the thermodynamic analysis of the solution. The quark-antiquark potential has been studied in section 4. We conclude in section 5 with a discussion.

\section{Black hole solution}

We will consider the Einstein gravity along with the Gauss Bonnet term as the higher derivative correction. For the matter content we include a Maxwell term  and a simplified action that is equivalent to the contribution from the  string cloud. 
Considering all the terms together the action is given by
\be\label{action}
S =  \int\limits_M d^{d+1}x \sqrt{-g} \big[ \frac{1}{\kappa^2} R - \Lambda + \alpha L_{GB} ]  + S_{MW} + S_{SC},
\ee
where $G$ is the gravitational constant, $g_{\mu\nu}$ is the metric tensor, $ \Lambda$ represents a cosmological constant term, $S_{MW}$ and $S_{SC}$ represent the Maxwell and the string cloud action respectively. The correction due to higher derivative terms are given by the Gauss-Bonnet term, $L_{GB}$\cite{Anninos:2008sj,Torii:2005nh}
\be
L_{GB} = R^2 - 4 R_{\mu\nu} R^{\mu\nu} + R_{\mu\nu\rho\sigma}R^{\mu\nu\rho\sigma}.
\ee

The action for the Maxwell term is given by \cite{Anninos:2008sj}
\be
S_{MW} = - \frac{1}{4 g^2} \int\limits_M d^{d+1}x \sqrt{-g} F_{\alpha\beta}F^{\alpha\beta} = - \int\limits_M d^{d+1}x \sqrt{-g} \frac{1}{4 g^2}  F^2 .
\ee
The energy momentum tensor for the Maxwell term is \cite{Anninos:2008sj}
\be\begin{split} T^{(EM)}_{\mu\nu} 
= \frac{1}{2g^2} \Big( F_{\mu\alpha} F_\nu^{~\alpha} - \frac{1}{4} g_{\mu\nu} F_{\alpha\beta}F^{\alpha\beta} \Big) .
\end{split}\ee

The Maxwell equation is given by
\be
\nabla_\alpha F^{\alpha\beta} =\frac{1}{\sqrt{-g}}\pa_\alpha (\sqrt{-g} F^{\alpha\beta}) = 0.
\ee

Assuming only the zeroeth component of the electromagnetic potential  $A_t$ to be non-zero, the Maxwell equation reduces to
\be
\pa_r ( r^{d-1} F^{rt})= 0
\ee
which implies
\be \label{emsoln}
F_{rt} = \frac{q}{ r^{d-1}} .
\ee
This leads to,
 \be \label{empot}
A_t = - \frac{q}{d-2}\Big( \frac{1}{r_H^{d-2}} - \frac{1}{ r^{d-2}} \Big) ,
\ee
where we have chosen the constant of integration in such a manner that $A_t$ vanishes at the horizon. Therefore the chemical potential, which can be obtained by taking asymptotic limit
\be\label{chempot}
\phi =  - \frac{q}{d-2} \frac{1}{r_H^{d-2}} .
\ee
\paragraph{}

The $\sigma$-model action for the string in canonical variables can be written as \cite{Chakrabortty:2011sp}
\be
S_{SC} = -T_2 \int d^2\xi \Big( \frac{1}{2} \sqrt{-\gamma} \gamma^{ij} \pa_i X^\mu \pa_j X^\nu g_{\mu\nu} 
\Big)
\ee
where $T_2$ is related to the string tension, $X^\mu$ and $g_{\mu\nu}$ ($\xi^i$ and $\gamma_{ij}$) are the space time (string worldsheet) coordinates and metrics respectively, $\gamma$ represents the determinant of $\gamma_{ij}$. 


We obtain the energy momentum tensor from the string cloud action as follows
\be
T^{(SC) {\mu\nu}} = - \frac{T_2}{2} \int d^2\xi \sqrt{-\gamma} \gamma^{ij} \pa_i X^\mu \pa_j X^\nu \frac{\delta^{(d+1)}(x-X)}{\sqrt{-g}}.
\ee
The solutions to these equations by making a choice of static gauge \cite{Chakrabortty:2011sp}
are
\be
X^t = \xi^0,\quad X^r=\xi^1,\quad X^\mu = y^\mu, \quad \text{for}\quad \mu\neq t, \mu \neq r
\ee
while other components are constant.

With the above choice of the static gauge the energy momentum tensor reduces to
\be
T^{(SC) tt} = - g^{tt} \frac{T_2}{2 r^{d-1}} \delta^{(d-1)}(x - X),\quad
T^{(SC) rr} =  - g^{rr} \frac{T_2}{2 r^{d-1}} \delta^{(d-1)}(x - X),
\ee
with the other components to be zero.

Considering a number of strings, following  \cite{Chakrabortty:2011sp} we will assume that the strings are uniformly distributed over the (d-1) directions and the density of the string cloud is represented by
\be \label{scdensity}
b(x) =  T_2 \sum\limits_{i}  \delta^{(d-1)}(x - X_i).
\ee
Since for negative $b$, the enrgy momentum tensor will not satisfy the weak and dominant energy condition \cite{Letelier:1979ej,Stachel:1980zr,Gibbons:2000hf}, we will assume $b \ge 0$.
Furthermore for the sake of simplicity, we will consider $b$ to be a constant. With this choice the energy momentum tensor of the string cloud turns out to be
\be
T^{(SC) tt} = - g^{tt} \frac{b}{2 r^{d-1}},\quad T^{(SC) rr} = - g^{rr} \frac{b}{2 r^{d-1}}
\ee

We can rewrite the on shell action for the string cloud using the string cloud density. Since we have incorporated the delta function in the density function $b(x)$ in (\ref{scdensity}), it can be written as
\be
S_{SC} = - \int d^{d+1}x ~ b
\ee

The equation of motion obtained from the action (\ref{action}) by varying the metric is given by
\be \label{eom}
\frac{1}{\kappa^2} ( R_{\mu\nu} - \frac{1}{2} g_{\mu\nu} R ) + \alpha H_{\mu\nu} + \frac{1}{2} g_{\mu\nu}\Lambda  = T^{(EM)}_{\mu\nu} + T^{(SC)}_{\mu\nu}
\ee
where $H_{\mu\nu}$ represents the contribution from the Gauss-Bonnet term,
\be
H_{\mu\nu} = 2[ R R_{\mu\nu} - 2 R_{\mu\alpha} R^{\alpha}_{~\nu} - 2 R^{\alpha\beta} R_{\mu\nu\alpha\beta} + R_\mu^{\alpha\beta\gamma} R_{\nu\alpha\beta\gamma} ] - \frac{1}{2} g_{\mu\nu} L_{GB}.
\ee

We consider the follwing ansatz for the metric
\be \label{ansatz}
ds^2 = - V(r) dt^2 + \frac{dr^2}{V(r)} + r^2 d\Omega_{d-1}^2
\ee
where $d\Omega_{d-1}^2$ is the round metric on a unit (d-1)-sphere. 

With the ansatz (\ref{ansatz}) the tt-component of the left hand side of (\ref{eom}) becomes
\be
\frac{1}{2} V(r) \Big[ -\Lambda + \frac{12\alpha V^\prime (V-1)}{r^3} + \frac{6(1-V) - 3 r V^\prime}{\kappa^2 r^2} \Big]
\ee
The energy momentum tensor of the EM field contributes
\be
T^{(EM)}_{tt} = \frac{1}{2} V F_{rt}^2 =   \frac{1}{2} V \frac{q^2}{ r^{2(d-1)}},
\ee
and the EM tensor of the string cloud contributes
\be
T^{(SC)}_{tt} =  V \frac{b}{2  r^{d-1}}.
\ee

The equation becomes
\be
\frac{1}{2} V(r) \Big[ -\Lambda + \frac{12\alpha V^\prime (V-1)}{r^3} + \frac{6(1-V) - 3 r V^\prime}{\kappa^2 r^2} \Big] =  \frac{1}{2} V \frac{q^2}{ r^{2(d-1)}} + V \frac{b}{2 r^{d-1}  }
\ee

This equation can be solved for $V$ and is given by
\be \label{V}
V(r) = 1 + \frac{r^2}{4\alpha \kappa^2} \Big[ 1 - \Big( 1 + \frac{2}{3} \alpha\kappa^4 \Lambda + \frac{4}{3(3-d)} \alpha \kappa^4 q^2  r^{2(1-d)} + \frac{8}{3(5-d)} \alpha \kappa^4 b r^{1-d} + \frac{\tilde{m}}{r^4}\Big)^{1/2} \Big]
\ee
where $\tilde{m}$ is the constant of integration. For convenience we will trade the constant $\tilde{m}$ for a constant $\mu$, where $\mu$ is related to the constant $\tilde{m}$ as
\be
\tilde{m} = - \frac{8}{3} \alpha \kappa^4 \mu .
\ee
Substituting $\mu$ for $\tilde{m}$ we get
\be \label{V}
V(r) = 1 + \frac{r^2}{4\alpha \kappa^2} \Big[ 1 - \Big( 1 + \frac{2}{3} \alpha\kappa^4 \Lambda + \frac{4}{3(3-d)} \alpha \kappa^4 q^2  r^{2(1-d)} + \frac{8}{3(5-d)} \alpha \kappa^4 b r^{1-d} - \frac{8}{3} \alpha \kappa^4 \frac{\mu}{r^4}\Big)^{1/2} \Big] .
\ee
We will use $d=4$ from now on.

The horizon radius corresponds to zero of $V(r)$
\be \label{horizon}
V(r_H)=0
\ee
which in turn determines $\mu$ in terms of $r_H$. We obtain,
\be \label{mass}
\mu =   \frac{ - 12 r_H^4 + \Lambda \kappa^2 r_H^6 - 24 \alpha\kappa^2 r_H^2 - 2 \kappa^2q^2  + 8 (b/2) \kappa^2 r_H^3 }{4 \kappa^2 r_H^2}
\ee

The temperature of the black hole can be obtained from the relation
\be\label{temperature}
T = \frac{1}{4\pi} V^\prime (r_H) ,
\ee
and is given by
\be\label{temperature1}
T = \frac{6 r_H^4 - \kappa^2 (q^2 +  b r_H^3 + \Lambda r_H^6) }{12 \pi (r_h^2 + 4 \kappa^2 \alpha ) r_H^3} .
\ee

For later use we also study the asymptotic expansion of $V(r)$. For large $r$, $V(r)$ reduces to
\be
V(r) \sim \frac{r^2}{l_0^2} + 1 - \frac{b_0}{2 r} - \frac{m_0}{r^2} + \frac{q_0^2}{r^4} .
\ee
As we observe  the expansion is similar to a charged AdS black hole once we introduce the following set of parameters.
\be\begin{split}\label{newpar}
\Lambda &= -\frac{12}{\kappa^2 l_0^2} + \frac{24\alpha}{l_0^4},\quad
K^2 = \frac{ 9 + 6 \alpha \Lambda \kappa^4}{\kappa^4}, \\
m_0 &= - \frac{\mu}{K} ,\quad
b_0 = \frac{b}{K},\quad
q_0^2 = \frac{q^2 K^2 + 6 \alpha b^2}{2 K^3} .
\end{split}\ee

The above equations impose a restriction on the range of the coefficient of the GB term $\alpha$ as $\alpha \leq \frac{l^2}{4 \kappa^2}$ once we choose $\Lambda = - \frac{6}{l^2}$. However, there are other restrictions on the GB coefficient $\alpha$. From the consideration of causality and unitarity it was found that the range of $\alpha$ is restricted to a dimension dependent interval \cite{Brigante:2007nu,Brigante:2008gz,Camanho:2009vw}. The interval was futher narrowed down due to constraints arising from causality \cite{Camanho:2014apa}.

\section{Thermodynamics}

We will begin our discussion of thermodynamics by deriving an expression for the free energy. The free energy can be obtained either using the first law of thermodynamics or from the on-shell Euclidean action and we will adopt the latter method here \cite{Chamblin:1999tk}. The on shell Euclidean action, as such, is divergent and so it requires appropriate subtraction. This can be done either by introducing counter terms \cite{Henningson:1998gx, Balasubramanian:1999re, Emparan:1999pm, Kofinas:2006hr, Astefanesei:2008wz} or subtracting the contribution of another geometry having identical asymptotic behaviour. We will use the second approach to evaluate the free energy.

The free energy for our black hole solution is obtained from the thermal partition function $Z$. The Euclidean metric is obtained by continuing to imaginary time and identifying it periodically with a period given by the inverse of Hawking
temperature. Using the saddle point approximation one can evaluate the partition function by
evaluating the classical action for the black hole metric.
So the free energy is given by\cite{Anninos:2008sj}
\be
F_1 = - k_B T \log Z = - k_B T S_E
\ee
where $k_B$ is the Boltzmann constant and $S_E$ is the Euclidean action. The Euclidean action is divergent and so we introduce a cut off at a large value $r = r_{\text{max}}$, subtract the Euclidean action of the appropriate reference background and then take the limit $r_{\text{max}} \rightarrow \infty$.  The free energy can be obtained by using the method of the counterterms \cite{Henningson:1998gx, Balasubramanian:1999re, Emparan:1999pm, Kofinas:2006hr, Astefanesei:2008wz} as well. This corresponds to the grand canonical ensemble, where the potential rather than the charge is kept fixed and so it represents the Gibbs free energy.

Before we compute the expression of free energy, one can observe that the expression of the on shell action can be substantially simplified by considering the trace of the equation of motion (\ref{eom}). We have
\be
g^{\mu\nu} H_{\mu\nu} = - \frac{d-3}{2} L_{GB} .
\ee
The trace of the energy momentum tensor coming from the Maxwell action,
\be
g^{\mu\nu} T^{(EM)}_{\mu\nu} = - \frac{d-3}{2} \frac{1}{4 g^2}F_{\alpha\beta}F^{\alpha\beta} = - \frac{d-3}{2} \frac{1}{4 g^2}F^2 .
\ee
Similarly the trace of the energy momentum tensor from the string cloud
\be
g^{\mu\nu} T^{(SC)}_{\mu\nu} = - \frac{2 b}{2 r^{d-1}}
\ee
Taking the trace of the equation of motion
\be
-\frac{1}{\kappa^2} \frac{d-1}{2} R + \frac{d+1}{2} \Lambda - \frac{d-3}{2} \alpha L_{GB} = - \frac{d-3}{2} \frac{1}{4 g^2}F^2 - \frac{2 b}{2 r^{d-1}}
\ee
we obtain
\be\label{relation}
\frac{1}{\kappa^2} R - \Lambda + \alpha L_{GB} - \frac{1}{4g^2} F^2   = - \frac{2}{d-3} \Big( \frac{1}{\kappa^2} R - 2 \Lambda - \frac{2 b}{2 r^{d-1}} \Big)
\ee

The free energy can be written as,
\be
- F_1\beta =  \int d^{d+1}x \sqrt{-g} \big[ \frac{1}{\kappa^2} R - \Lambda + \alpha L_{GB} - \frac{1}{4g^2} F^2  \big]  {-} \int dt dr b \Sigma_{d-1}
\ee
where we have used $\Sigma_{d-1}$ is the volume of the space transverse to the t and r direction and $\beta$ is the inverse temperature. Using the relation given in (\ref{relation}), we obtain
 \be
- F_1\beta =  \int d^{d+1}x \sqrt{-g} \big[  - \frac{2}{d-3} \big( \frac{1}{\kappa^2} R - 2 \Lambda -  \frac{2 b}{ 2 r^{d-1}} \big)\big] {-} \int dt dr b \Sigma_{d-1}
\ee

In order to regularize the divergences ensuing from the asymptotic limit, we need to subtract away the free energy of the reference background, which  
we choose to be the neutral extremal black hole obtained by setting $T = q =0$ \cite{Anninos:2008sj}. From the expression of the temperature given in (\ref{temperature1}) we obtain
\be \label{rveqn}
r_v^3 - \frac{6}{\Lambda \kappa^2}r_v + \frac{2 b}{2 \Lambda}=0 ,
\ee
and the radius of the horizon of the reference black hole $r_v$ is a root of the above cubic equation. As we observe, that in absence of the string cloud $r_v=0$ would have been a solution, while its presence leads to a finite horizon radius.

We can check the nature of $r_v$ using Cardano's formula.  Any cubic equation can be expressed as a depressed cubic, written as
\be
t^3 + p_1 t + p_0 =0,
\ee
and according to that formula, the discriminant of a depressed cubic is given by
\be
\triangle = - ( 4p_1^3 + 27 p_0^2).
\ee
For $\triangle \ge 0$ all the three roots are real and for $\triangle \le 0$ it has one real and two non-real roots.
In the present case,
\be
\triangle = - \Big( - 4 \frac{6^3}{\Lambda^3 \kappa^6} + 108 \frac{b^2}{4 \Lambda^2} \Big).
\ee
Considering the fact that the present case has a negative cosmological constant, $\Lambda \le 0$  we find that $\triangle \le 0$ and it has only one real root.
The expression of the real root can be obtained from Cardano's formula
\be
t = \Big( - \frac{p_0}{2} + \sqrt{\frac{p_0^2}{4} + \frac{p_1^3}{27}} \Big)^{1/3} + \Big( - \frac{p_0}{2} - \sqrt{\frac{p_0^2}{4} + \frac{p_1^3}{27}} \Big)^{1/3}.
\ee
Substituting the coefficients in the depressed cubic in the above formula, we obtain,
\be\label{horizon-ref}
r_v = \Big( - \frac{b}{2 \Lambda} + \sqrt{\frac{b^2}{4  \Lambda^2} - \frac{6^3}{27 \Lambda^3 \kappa^6}} \Big)^{1/3} +  \Big( - \frac{b}{2 \Lambda} - \sqrt{\frac{b^2}{4 \Lambda^2} - \frac{6^3}{27 \Lambda^3 \kappa^6}} \Big)^{1/3}.
\ee
Since in our case $\Lambda \le 0$, one can see that $r_v \ge 0$ for any choice of the parameters and so the neutral extremal limit of the black hole can serve as a reference backround has a finite positive horizon radius.
The expression for the mass parameter of the reference black hole as obtained from (\ref{mass}) is given by
\be \label{mass-ref}
\mu_{\text{vac}} = -\frac{3}{2\kappa^2} r_v^2 + \frac{3b}{4} r_v - 6\alpha
\ee

 Using the neutral extremal limit of the black hole as the reference backround and making the necessary subtraction leads to the following expression for the free energy,
\be\begin{split}
- \frac{F_1}{\Sigma_{d-1}} &=  \int\limits_{r_H}^{r_{\text{max}}} dr r^{d-1} \big[  - \frac{2}{d-3} \big( \frac{1}{\kappa^2} R - 2 \Lambda -  \frac{2 b}{2 r^{d-1}} \big)  {-} \frac{ b}{r^{d-1}}  \big] \\
&- \sqrt{\frac{V(r_{\text{max}})}{V_0 (r_{\text{max}})}}\int\limits_{r_v}^{r_{\text{max}}} dr r^{d-1} \big[  - \frac{2}{d-3} \big( \frac{1}{\kappa^2} R - 2 \Lambda -  \frac{2 b}{2 r^{d-1}} \big) {-} \frac{ b}{r^{d-1}}  \big]
\end{split}\ee

For our purpose we set $d=4$ and the expression of the free energy reduces to
%
\be\begin{split}
- \frac{F_1}{\Sigma_3} &=  \int\limits_{r_H}^{r_{\text{max}}} dr \big[  \frac{2}{\kappa^2} (- r^3 R) + 4 \Lambda r^3   +   b  \big] \\
&- \sqrt{\frac{V(r_{\text{max}})}{V_0 (r_{\text{max}})}}\int\limits_{r_v}^{r_{\text{max}}} dr  \big[  \frac{2}{\kappa^2} (- r^3 R) + 4 \Lambda r^3   +   b  \big] .
\end{split}\ee
In order to simplify the above expression we can use the following relation \cite{Anninos:2008sj},
\be
R = -\frac{1}{r^{d-1} }\Big(r^{d-1} V(r)\Big)^{\prime\prime} + \frac{2(d-1)}{r^2} ,
\ee
which for $d=4$ reduces to
\be
R = -\frac{1}{r^3 }\Big(r^3 V(r)\Big)^{\prime\prime} + \frac{6}{r^2} .
\ee

Substituting
%
%
and carrying out the
integration we obtain
\be\begin{split}\label{fe1}
- \frac{F_1}{\Sigma_3} &= \Big[ \frac{2}{\kappa^2}
( r^3 V(r))^\prime   -\frac{6}{\kappa^2} r^2  +  \Lambda r^4   +  b r \Big]_{r_H}^{r_{\text{max}}}  \\
& - \sqrt{\frac{V(r_{\text{max}} )}{V_0 (r_{\text{max}})}} \Big[\frac{2}{\kappa^2}
( r^3 V(r))^\prime   -\frac{6}{\kappa^2} r^2  +  \Lambda r^4   +  b r \Big]_{r_v}^{r_{\text{max}}} .
\end{split}\ee

The prefactor in front of the contribution from the reference black hole in (\ref{fe1}), which has been introduced to match up the asymptotic behaviour can be expanded as
\be
 \sqrt{\frac{V(r_{\text{max}} )}{V_0 (r_{\text{max}})}}  = 1 - \frac{(m_0-m_0^{\text{vac}}) l_0^2}{2 r_{\text{max}}^4} + {\mathcal{O}}(\frac{1}{r_{\text{max}}^5})
\ee

We choose the limit $r_{\text{max}} \rightarrow \infty$ and further use the relation in (\ref{horizon}) and (\ref{temperature})  to reduce the first terms on the RHS of each line in the above equation
\be \lim_{r\rightarrow r_H} ( r^3 V(r))^\prime = 4\pi T r_H^3 ,\quad \lim_{r\rightarrow r_{\text{vac}}} ( r^3 V(r))^\prime = 0 \ee
since the temperature of the reference black hole is zero. After making use of these relations, the free energy  reduces to
\be\begin{split}\label{fe2} - \frac{F_1}{\Sigma_3} &= (\mu - \mu_{\text{vac}}) - \big[\frac{8 \pi T}{\kappa^2} r_H^3    -\frac{6}{\kappa^2} r_H^2  +  \Lambda r_H^4   +   b r_H \big] \\
&+ \big[  -\frac{6}{\kappa^2} r_v^2  +  \Lambda r_v^4   +  b r_v\big]
\end{split}\ee

Therefore we can write down
\be\begin{split}\label{felo3}
 \frac{F_1}{\Sigma_3} &= - \big[ \mu  - \frac{8 \pi T}{\kappa^2} r_H^3   + \frac{6}{\kappa^2} r_H^2  -  \Lambda r_H^4   -   b r_H \big] - \frac{ F_{\text{vac}}}{{\Sigma_3}} \\
  \frac{F_{\text{vac}}}{\Sigma_3} &= -  \big[ \mu_{\text{vac}}  + \frac{6}{\kappa^2} r_v^2  -  \Lambda r_v^4   -   b r_v \big]
\end{split}\ee
Since this is a grand canonical ensemble, we consider the electric potential instead of the charge to be fixed and is given by (\ref{chempot})
\be
\phi =  - \frac{q}{d-2} \frac{1}{r_H^{d-2}}.
\ee

We obtain the expression of $I_1 = \beta F$ following \cite{Chamblin:1999tk} after the necessary subtraction as 
\be\begin{split}
 I_1&= - \frac{2\pi \Sigma_3}{- 6 r_H^3 \kappa^4 + l^2 (-6\kappa^2 + \kappa^4 b + 4 \kappa^4 r_H \phi^2)}
 \times \Bigg( 
3 \kappa^2 \Big[ r_H^6 - 9 r_H^2 r_v^4 + 36 \alpha \kappa^2( r_H^4 - r_v^4) \Big]\ \\
&+ 2 l^2 \Big[ 2 \kappa^2 b r_H^3  -36 \alpha \kappa^2 r_v^2 +  ( 2 \kappa \phi^2 -3)r_H^4 - 3 r_H^2 (3 r_v^2 -12 \alpha \kappa^2 + 8 \alpha \kappa^4 \phi^2) \Big]
\Bigg),
\end{split}\ee
where we have used $\Lambda = - \frac{6}{l^2}$.
One can obtain various thermodynamic quantities as follows. The Gibbs potential for grand canonical ensemble can be written as
\be
F = I_1/\beta = E - TS - \phi ~ Q
\ee

So the charge can be obtained as
\be
Q = -\frac{1}{\beta}\left( \frac{\partial I_1}{\partial \phi} \right)_\beta = -2 q \Sigma_3.
\ee
Similarly, the entropy is given by 
\be
S = \beta \left(\frac{\partial I_1}{\partial\beta}\right)_\phi - I_1 = 4 \pi r_H \left(12 \alpha + \frac{r_H^2}{\kappa^2}\right) \Sigma_3
\ee
The energy is given by
\be
E = \left(\frac{\partial I_1}{\partial\beta}\right)_\phi - \frac{\phi}{\beta} \left(\frac{\partial I_1}{\partial\phi}\right)_\beta
= \frac{1}{2} \Sigma_3 \left(\frac{q^2}{r_H^2} -2 b r_H + \frac{3 r_H^4}{l^2} + \frac{9 r_v^4}{l^2}+\frac{6 r_H^2}{\kappa ^2}+\frac{6 r_v^2}{\kappa ^2}\right) 
\ee

In addition we can introduce
\be M = - ( \mu - \mu_v) 
\ee  
which turns out to be
\be M = \frac{1}{2} \left( \frac{q^2}{r_H^2} + \frac{3 r_H^4}{l^2} - 2 b (r_H-r_v) - \frac{3 r_v^4}{l^2} + \frac{6 r_H^2}{\kappa ^2} - \frac{6 r_v^2}{\kappa ^2}\right)
\ee

A straightforward computations shows that
\be
\frac{E}{\Sigma_3} - M =  r_v  \left(- b r_v + 6 r_v ( \frac{r_v^2}{l^2} + \frac{1}{\kappa^2})\right)
\ee
which vanishes by using the equation (\ref{rveqn}) satisfied by $r_v$ and thus establishes the fact that
\be E = M \Sigma_3 .
\ee

From the above expressions we obtain
\be
\frac{\partial E}{\partial Q} = \phi ,\quad \frac{\partial E}{\partial S} = T
\ee
leading to the first law of thermodynamics $\text{$\delta$E} = \text{T $\delta$S} + \text{$\phi$ $\delta$Q}$

From the above expressions we can obtain the specific heat given by,
\be
C_v = T\left( \frac{\partial S}{\partial T} \right) .
\ee
The specific heat plays an important role to analyse the stability of the black hole solutions and for the present solution, it can be written as,
\be
C_v  =  \frac{12 \pi\sigma_3 \left( r_H^2 + 4 \alpha\kappa^2 \right)^2 \left( b \kappa ^2 + \kappa ^2 \Lambda r_H^3 + r_H \left(  -6 + 4 \kappa ^2 \phi ^2 \right) \right) }{ -2 b \kappa ^4 r_H +\kappa ^4 \Lambda r_H^4 + 2 \kappa ^2 r_H^2 \left(3 + 6 \alpha  \kappa ^4 \Lambda -2 \kappa ^2 \phi ^2 \right)+ 8 \alpha  \kappa ^4 \left(-3 + 2 \kappa ^2 \phi ^2\right)}
\ee

\begin{figure}[h]
\begin{center}
\begin{psfrags}
\mbox{\subfigure[]{\includegraphics[width=7.5 cm]{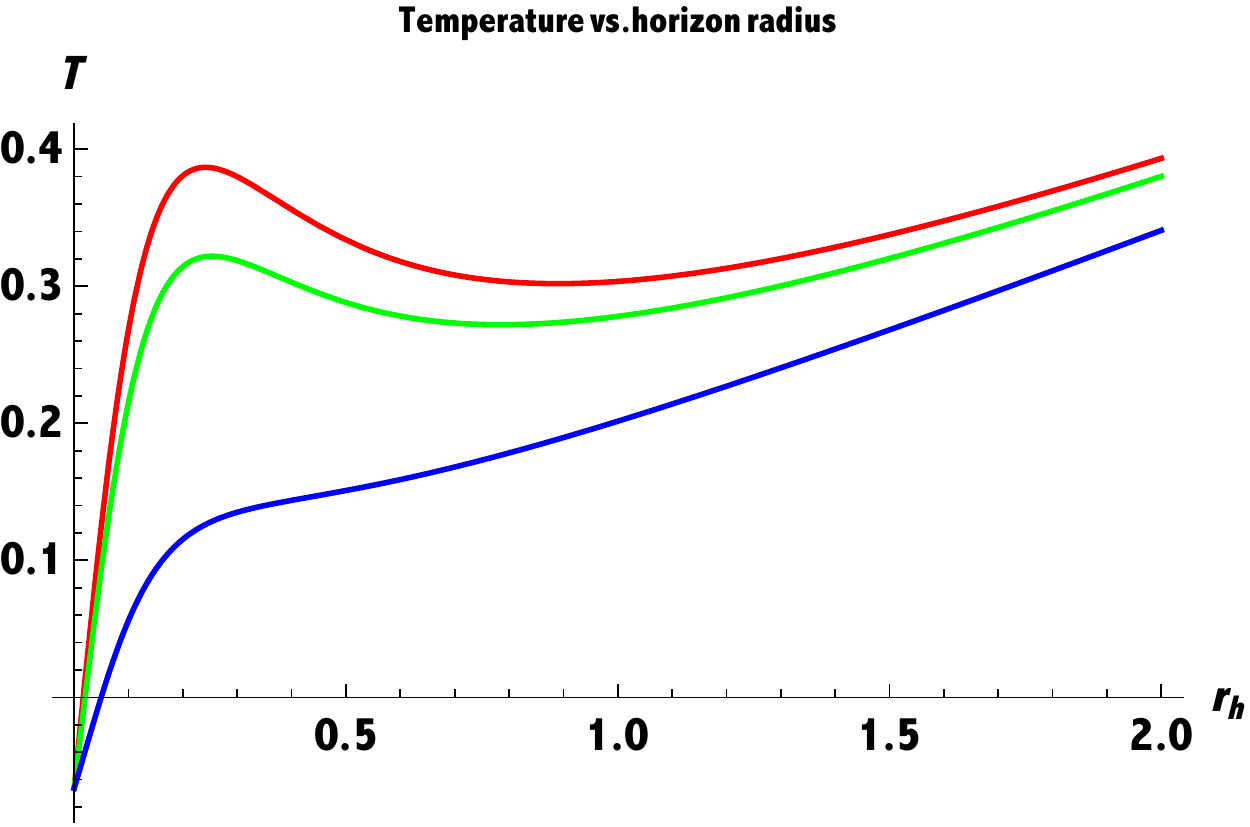}}
\quad
\subfigure[]{\includegraphics[width=7.5 cm]{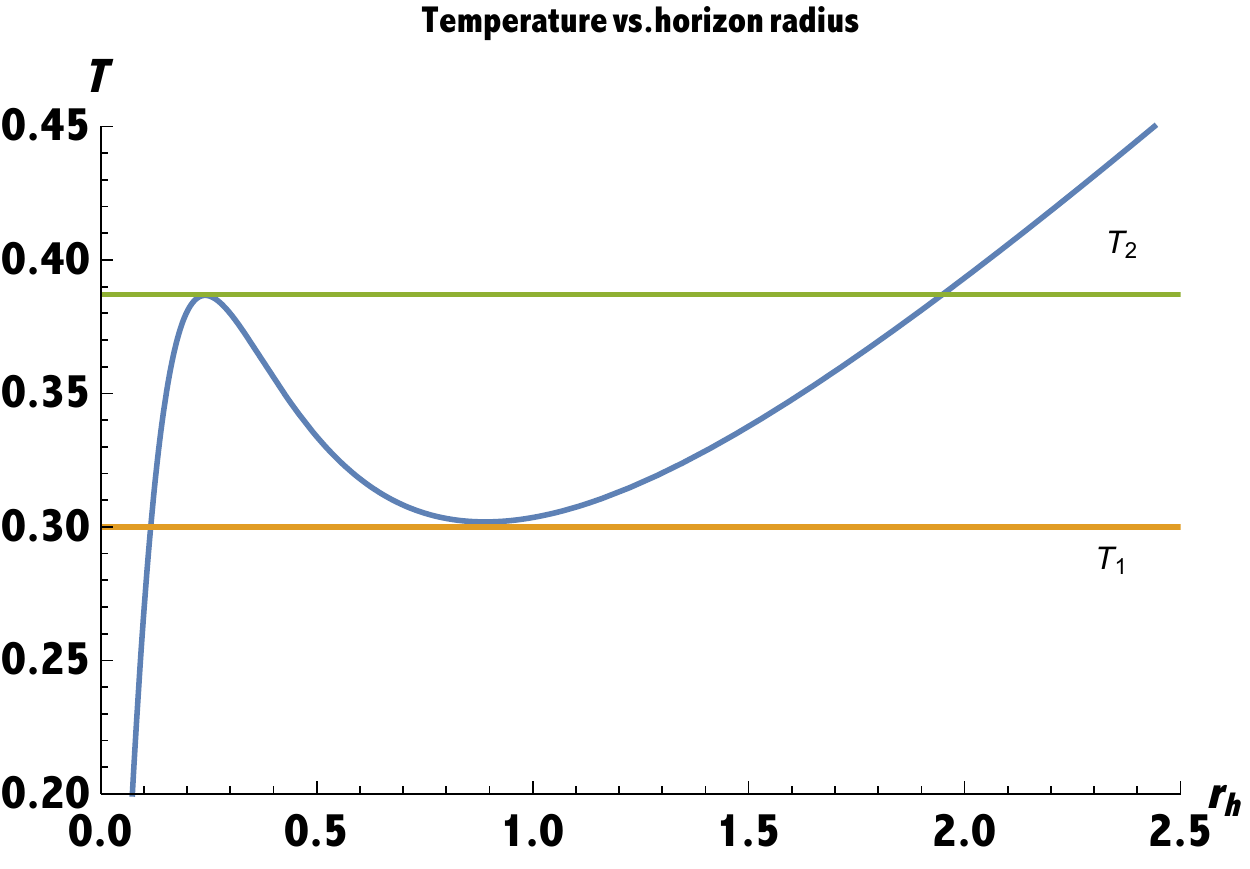}}
}
\end{psfrags}
\caption{Temparature is plotted against horizon radius for $\alpha=.01$, $ b=0.1$. (a) Three lines correspond to $\phi =$ 0 (red), 0.5 (green), 1(blue) respectively. (b)  Exhibiting the two transition temperatures for $\phi=0$.}
\label{Trh}
\end{center}
\end{figure}

In what follows, we will discuss relevant thermodynamic quantities associated with this black hole solution. We will begin with the temperature
and plotted the temperature vs. horizon radius for several values of the chemical potential for  certain fixed values of $\alpha$ and $b$ in Fig.\ref{Trh}. The plot for zero chemical potential is given in Fig.\ref{Trh}(b), where one observe that for low enough temperature, there exists only one solution for the black hole. As  the temperature increases beyond certain critical temperature $T_1$, there exists three black hole solutions with different radii. According to the comparative size if the radii, they are termed as small, intermediate and large black hole. As the temperature increases further, the intermediate black hole radius approaches that of the small black hole. Beyond certain temperature $T_2$, the intermediate black hole merges with the small black hole and they cease to exists. For $T\ge T_2$, only the large black hole exists.
Fig.\ref{Trh}(a) shows the change in the behaviour of the temperature vs. horizon radius plot as the chemical potential changes. One can see from that, as the chemical potential increases, the range of temperature in which the three black holes exist narrows down and for the chemical potential beyond certain critical value only one black hole persists.

\begin{figure}[h]
\begin{center}
\begin{psfrags}
\mbox{\subfigure[]{\includegraphics[width=7.5 cm]{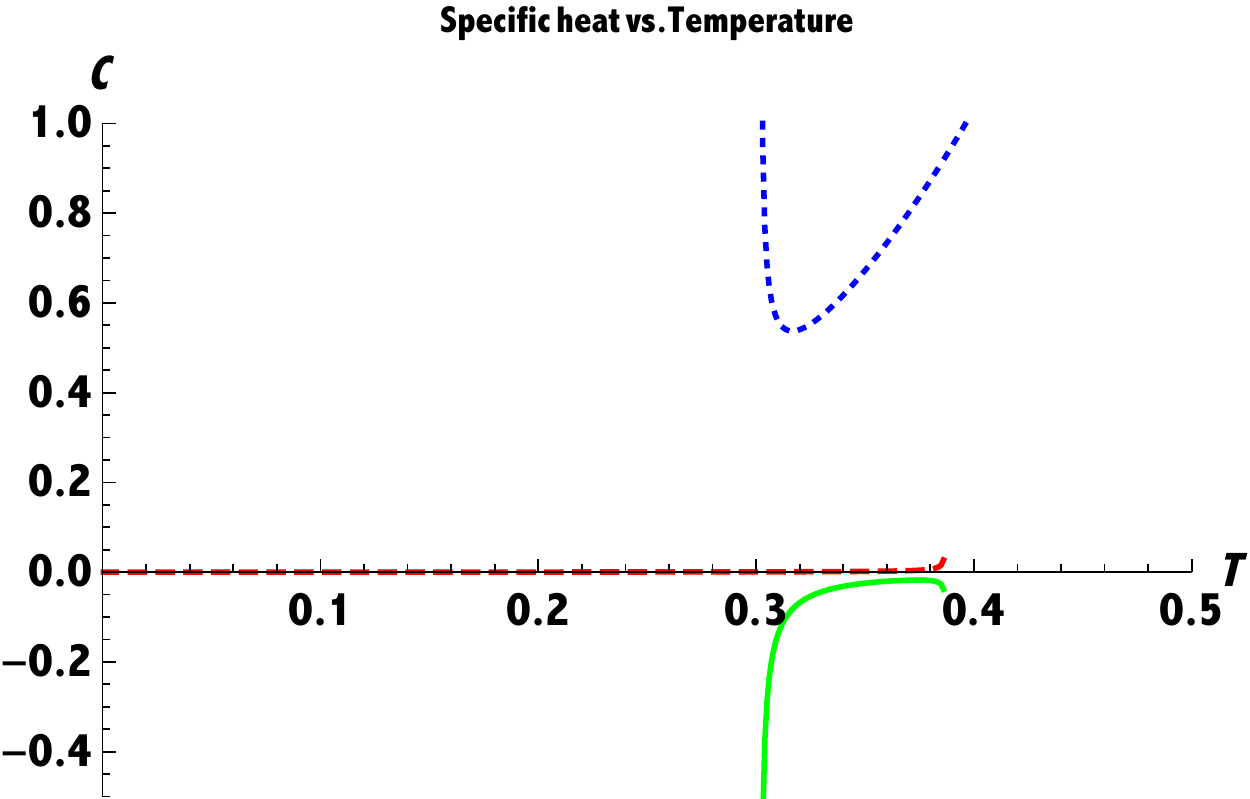}}
\quad
\subfigure[]{\includegraphics[width=7.5 cm]{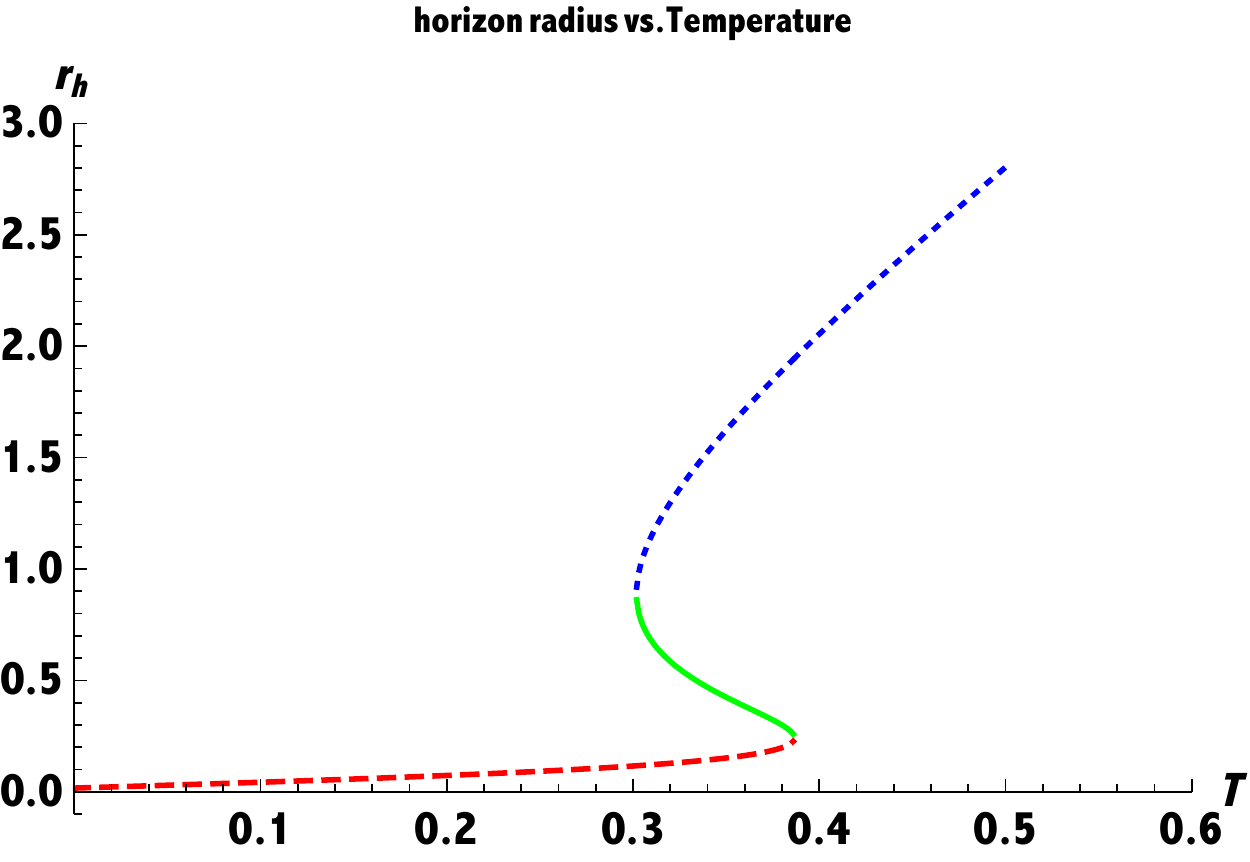}}
}
\end{psfrags}
\caption{(a) Specific heat and (b) Horizon radius are plotted against temparature for $\alpha=.01$, $ b=0.1$ and  $\phi=0$: Red green and blue segments correspond to the small, medium and large black hole repectively}
\label{spht0}
\end{center}
\end{figure}

In Fig.\ref{spht0}(a) the specific heat is plotted against the temperature for zero chemical potential. In oreder to identify the different black holes, we have also plotted the horizon radius vs. temperature in (b). As one can see the specific heat steadily increases with temperature for the small black hole and it remains positive indicating the fact that the small black hole is a stable configuration. The specific heat of the large black hole is greater and is also positive. On the other hand the specific heat of the intermediate black hole remains negative throughout the range signalling its instability. 

\begin{figure}[h]
\begin{center}
\begin{psfrags}
\mbox{\subfigure[]{\includegraphics[width=7.5 cm]{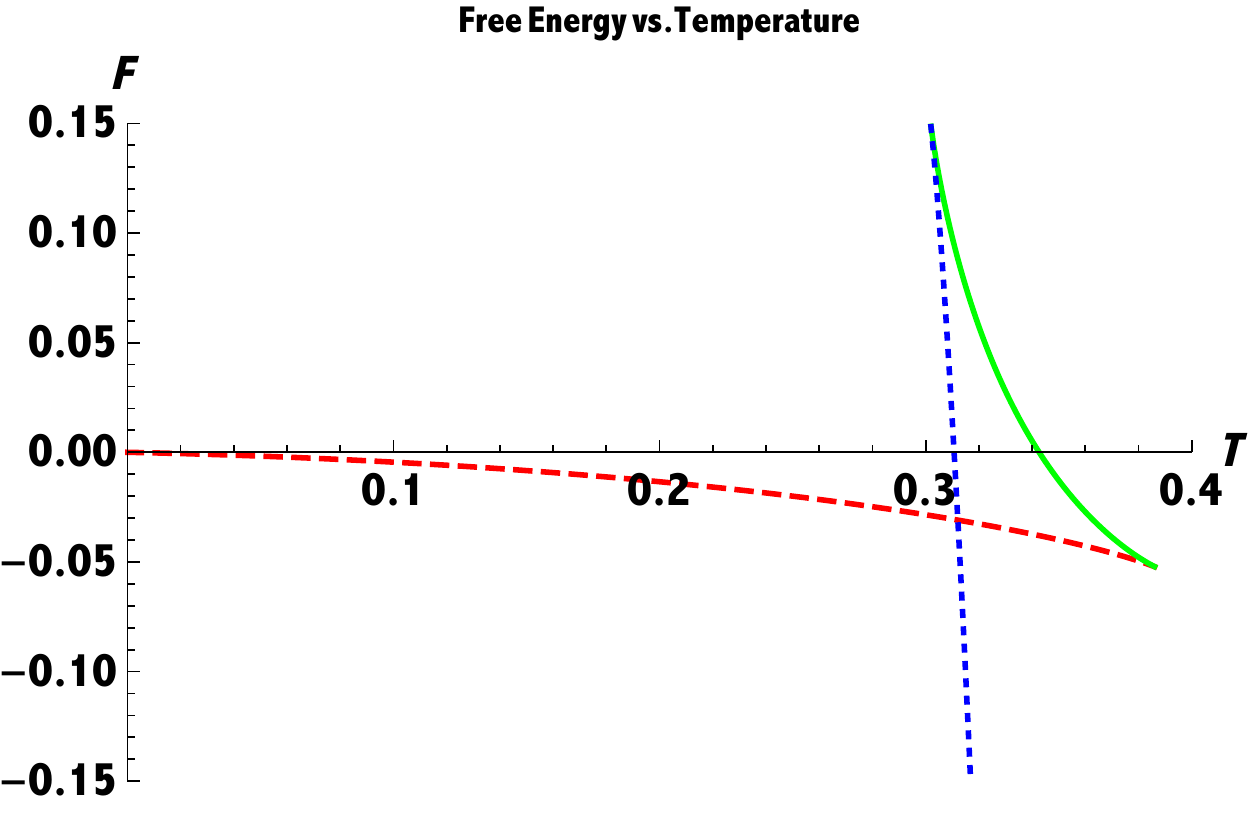}}
\quad
\subfigure[]{\includegraphics[width=7.5 cm]{HorizonRadiusVST1}}
}
\end{psfrags}
\caption{(a) Free Energy and (b) Horizon radius are plotted against temparature for $\alpha=.01$, $ b=0.1$ and  $\phi=0$: Red green and blue segments correspond to the small, medium and large black hole repectively}
\label{freeenergy}
\end{center}
\end{figure}

The thermodynamic free energy of the three different black hole are plotted in Fig.\ref{freeenergy}(a). As one can observe, the small black hole remains thermodynamically stable over a range of temperature, while beyond certain temperature the large black hole is the stable configuration. The medium black hole always remain unstable and the phase transition involves are of first order. 

\begin{figure}[h]
\begin{center}
\begin{psfrags}
\mbox{\subfigure[]{\includegraphics[width=7.5 cm]{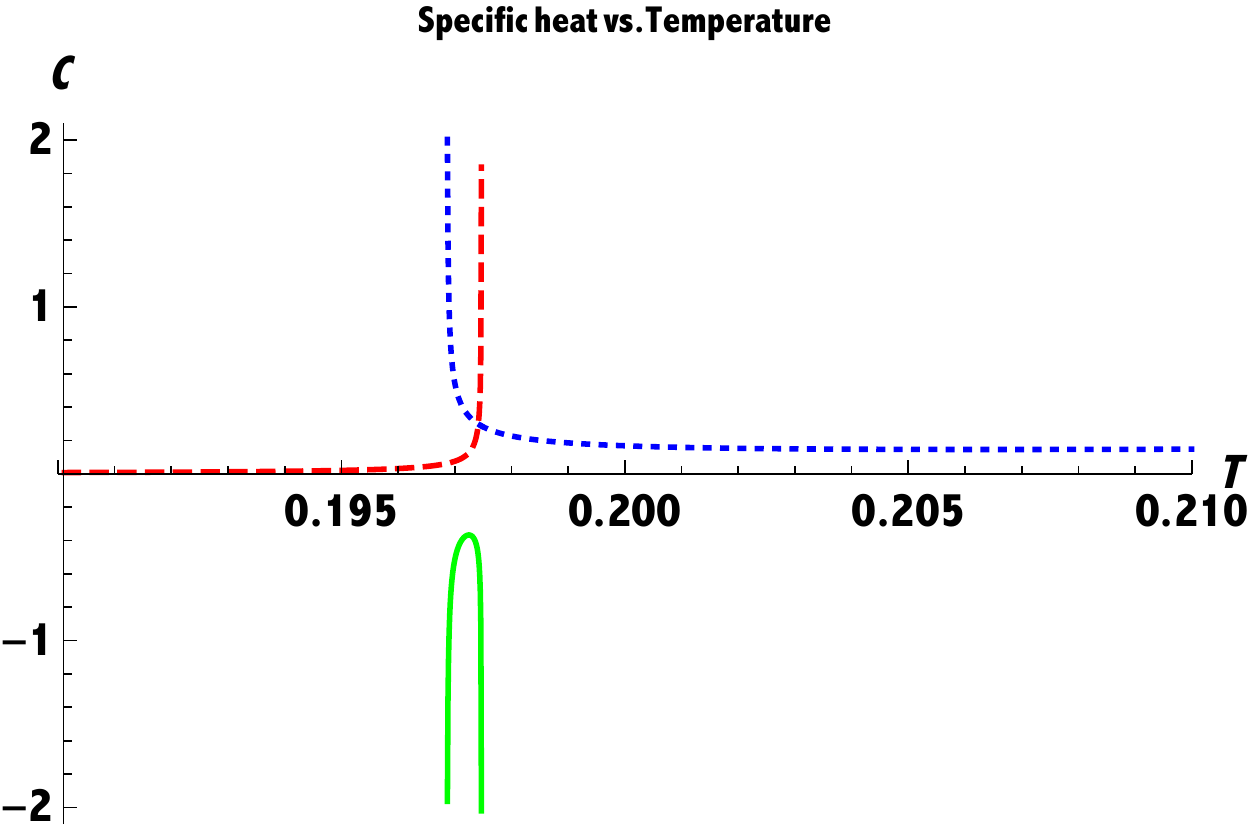}}
\quad
\subfigure[]{\includegraphics[width=7.5 cm]{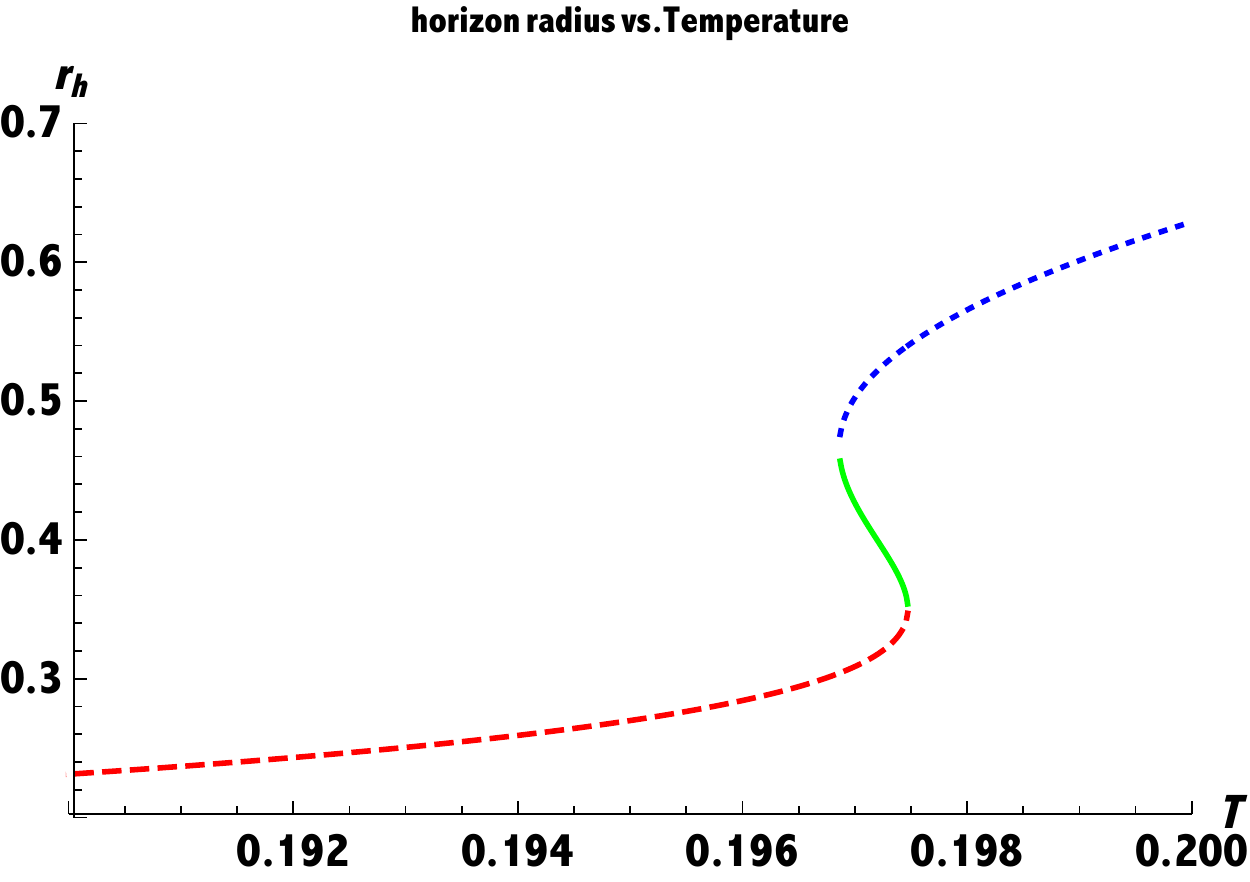}}
}
\end{psfrags}
\caption{(a) Specific heat and (b) Horizon radius are plotted against temparature for $\alpha=.01$, $ b=0.1$ and  $\phi=0.78$: Red green and blue segments correspond to the small, medium and large black hole repectively}
\label{spht2}
\end{center}
\end{figure}

\begin{figure}[h]
\begin{center}
\begin{psfrags}
\mbox{\subfigure[]{\includegraphics[width=7.5 cm]{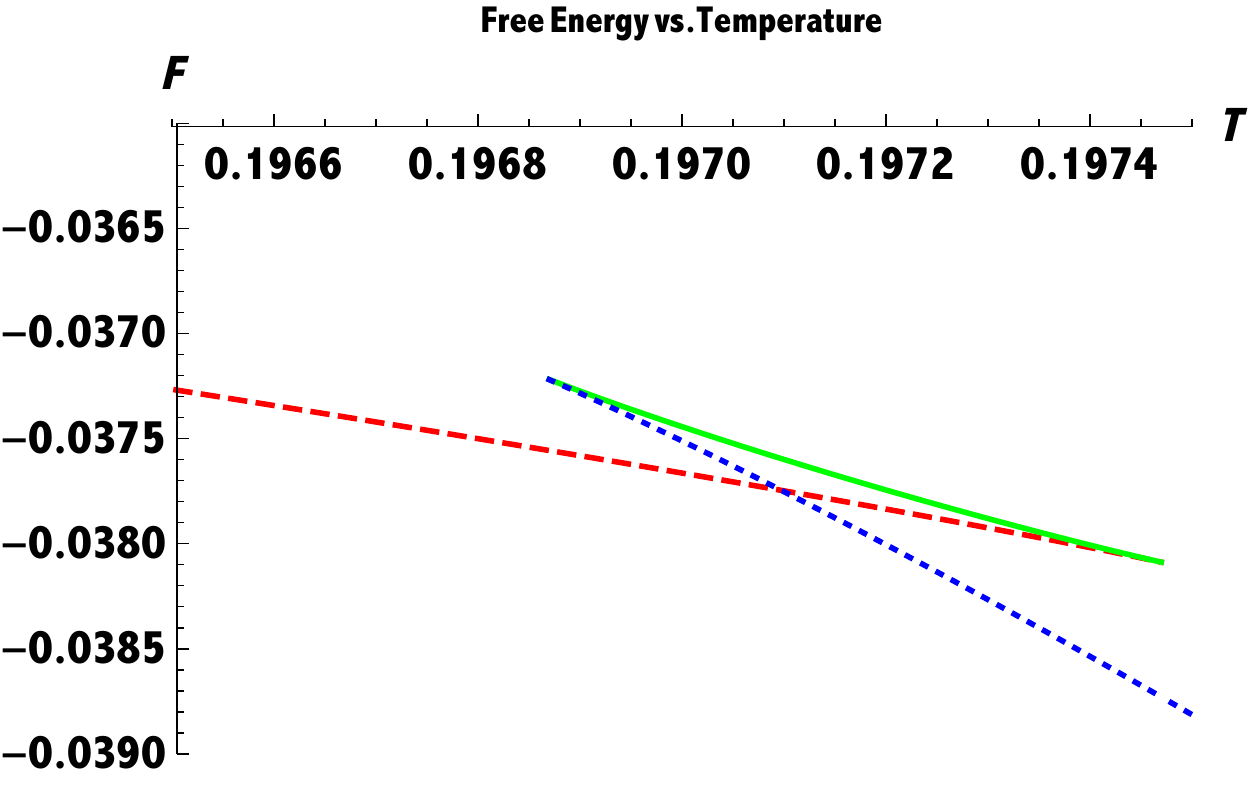}}
\quad
\subfigure[]{\includegraphics[width=7.5 cm]{HorizonRadiusVST2}}
}
\end{psfrags}
\caption{(a) Free Energy and (b) Horizon radius are plotted against temparature for $\alpha=.01$, $ b=0.1$ and  $\phi=0.78$: Red green and blue segments correspond to the small, medium and large black hole repectively}
\label{freeenergy2}
\end{center}
\end{figure}

As we have already mentioned that for large enough chemical potential the range of temperature in which, all the three blck holes exist bacoming narrower and at a critical value, that range vanishes leaving only a single black hole for any temperature. In Fig.\ref{spht2} and Fig.\ref{freeenergy2} we have plotted the specific heat and the free energy respectively with a value of the chemical potential close to the critical value. As evident from these plots, the specific heat exhibits a steep increase, once the region of the three holes is approached. The specific heat associated with the medium black hole remains negative. The triangle in the free enrgy curve becomes quite thin showing that this is on the verge of culminating to a single blck hole solution.

\section{Quark-antiquark distance}\label{gaugetheory}

One interesting phenomenon that we can explore using this black hole is to study quark-antiquark ($q\bar{q}$) distance  in a $q\bar{q}$ bound state in the boundary theory. Essentially one considers  string configurations having its two endpoints on the boundary at a distance $L$, and elongated in the bulk, which could be a  straight string from boundary to horizon or a U-shaped string where the string is hanging from the boundary, with its tip located at $u_0$, but not touching the horizon.   
 For every $L$ there are two U-shaped string configurations with tip $u_0$. One is nearer to the horizon and farther from the horizon. As $L$ increases, these two configurations approach each other and they merge at a maximum value $L=L_s$, when the quark-antiquark distance reaches the screening length.

Here we consider a probe open string in the black hole spacetime and obtain the screening length. The Nambu-Goto world sheet action is
\begin{equation}{\label{ng}}
S_{NG} = - \frac{1}{2 \pi \alpha^\prime } \int d^2\sigma \sqrt{- det (h_{\beta \gamma})},
\end{equation}
with string tension to be  $\frac{1}{2 \pi \alpha^\prime}$, where $\alpha^\prime$ is  related to $\lambda$, the `t Hooft coupling in the dual SYM gauge theory \cite{Ewerz:2016zsx},
\begin{equation}
\sqrt{\lambda} = \frac{l^2}{\alpha^\prime}.
\end{equation}
Though we do not have a precise description of the dual boundary theory, we can expect that the $\lambda$ will play an analogous role. We will assume that the above Nambu-Goto action is a valid action for our problem and corrections ensuing from coupling to the further bulk fields will not qualitatively modify the essential result.

 We trade $u$ for $r$, where $r= \frac{l^2}{u}$,  the boundary and the horizon occurs at u = 0 and $u_h$ respectively and the metric tensor (\ref{ansatz}) becomes,
 \begin{equation}\label{dualgrav}
ds^2= f(u)\big[-h(u)dt^2 + dx^2 + dy^2 +  dz^2 + {du^2\over h(u)}\big],
\end{equation}
\begin{equation}
{\rm where}, f(u)= {l^2\over u^2} \quad 
{\rm and}\quad 
h(u)= \frac{u^2}{l^2} + \frac{1}{4 \alpha} \Big( 1 - \sqrt{1 - 4 \alpha - \frac{8 \alpha \mu u^4}{3 l^4}  - \frac{4 \alpha q^2 u^6}{3 l^6} +  \frac{8 b \alpha u^3}{3 l^3}}\Big).\nonumber
\end{equation}
The horizon radius, $u_h$ is given by,
\begin{eqnarray}
 h(u_{h})= \frac{u_h^2}{l^2} + \frac{1}{4 \alpha}\Big(1-\sqrt{1-4\alpha - \frac{8 \alpha \mu u_h^4}{3 l^4}  - \frac{4 \alpha q^2 u_h^6}{3 l^6} + \frac{8 b \alpha u_h^3}{3 l^3}}\Big) = 0.
\label{hor}
\end{eqnarray}
We will consider, the end points of the open string corresponding to $Q\bar Q$ pair are located at $x = \pm \frac{L}{2}$ respectively and is elongated in the bulk space time with the symmetry around $x = 0$.

 We choose the static gauge  $\sigma^0 = t,\,\, \sigma^1 = x$, with the induced metric $h_{\beta \gamma} = \partial_\beta X^\mu\partial_\gamma X^\nu g_{\mu\nu}$ becomes,
\begin{equation}
ds^2 = f(u)\big[-h(u)dt^2 + \big\{1 + \frac{u'^2}{h(u)}\big\}dx^2\big].
\end{equation}
After $t$ integration is carried out, the Nambu-Goto action becomes
\begin{equation}\label{ng2}
S_{NG} = - \frac{{\mathcal T}}{2 \pi \alpha^\prime } \int\limits_{-L/2}^{L/2} dx f(u) \sqrt{h(u) \left( 1 +\frac{ {u^\prime}^2}{h(u)} \right)}.
\end{equation}
The equations of motion for $u(x)$, ensuing from the above action is given by
\begin{equation}\label{ng3}
u^\prime (x) = - \sqrt{ h(u)  \left( \frac{f(u)^2 h(u)}{f(u_0)^2 h(u_0)} - 1 \right)},
\end{equation}
implying the string is extended towards the horizon up to a turning point $u=u_0$ and goes back to the boundary in a symmetric manner with $u(0)=u_0$.
The distance between the two end points of the string can be obtained as
\begin{equation}\label{qqseparation}
L = \int\limits_{- L/2}^{L/2} dx = 2 \int\limits_0^{u_0} \frac{du}{u'} =2 \int\limits_0^{u_0}du \left[ h(u)
\left(  \frac{f(u)^2 h(u)}{f(u_0)^2 h(u_0)} -1 \right) \right]^{-\frac{1}{2}},
\end{equation}
where $u_0$ is the upper bound on the location of the tip of the string that is extended towards horizon.

\begin{figure}[h]
\begin{center}
	\mbox{
	\subfigure[]{\includegraphics[width=7.5cm]{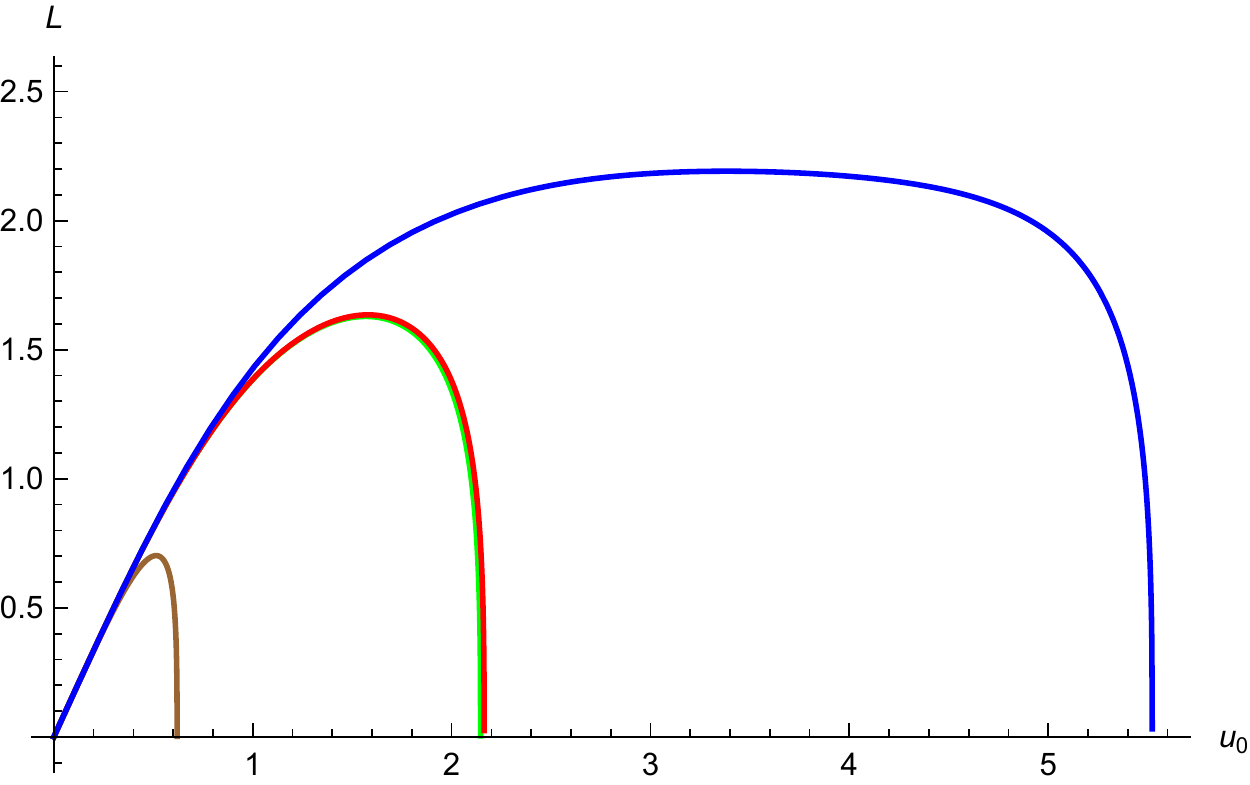}}
\quad
	\subfigure[]{\includegraphics[width=7.5cm]{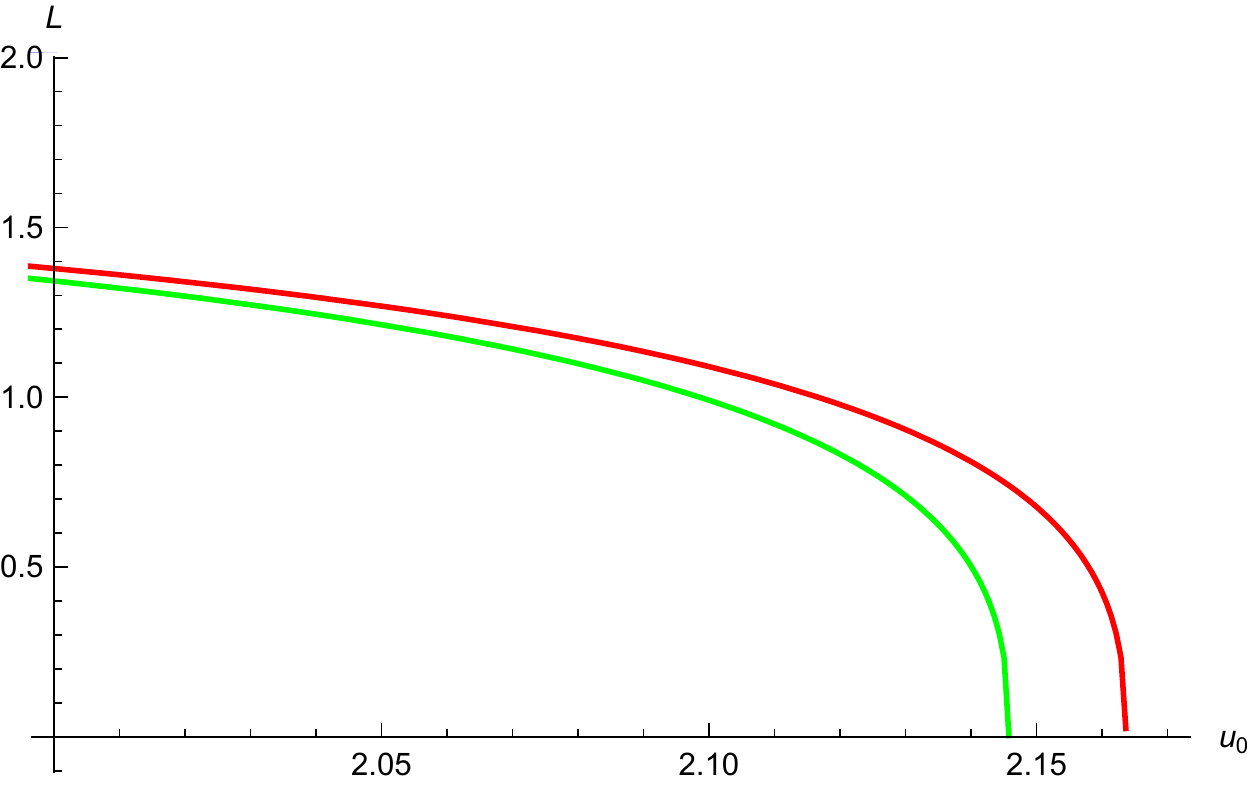}}}
\end{center}
\caption{Q$\bar Q$ -distance vs. $u_0$: $\alpha=.01, b = .01$, $q=.03$ and $T=.35$  black holes with radius , small (Blue), medium (Red),  large (Brown) on left. In addition, for comparison we have also given one plot for $L$ with $q=0$ for the large black hole. Since it is almost coincident with the medium black hole, we have given a magnified plot on right}
\label{distance}
\end{figure}

 We have plotted the quark-antiquark separation $L$  with respect to $u_0$, the upper bound on the location of the tip of the string extended towards the horizon in figure \ref{distance}. One can observe from the figure, as the tip of the string ( $u_0$) gets closer to horizon,  quark-antiquak separation keeps on increasing and reaches maximum when $u_0$ gets extremely near to the horizon. $L$ drops down to zero as $u_0$ touches the horizon. It shows breaking of the U-shaped string to two parallel straight string configuration. If the latter configuration is interpreted as unbound state of $Q\bar Q $ pair,  it implies that confinement is not present in the dual theory living on the boundary. We have studied till very low temperature but the feature remains the same. This happens in case of the small, medium as well as the large black hole. 
 
 The present study is for the ensemble where charge $q$ is fixed. A similar analysis with a fixed chemical potential $\phi$ can also be done and it is quite straightforward. We have studied the $q\bar{q}$ distance in that ensemble for different values of chemical potentials. Once again it does not show confinement. So we are not included those plots here.
\section{Discussion}

Higher derivative corrections are common feature of the UV completion of Einstein gravity and it also provides $\frac{1}{\sqrt{\lambda}}$ corrections in the gauge theory. In view of that, the present work deals with black hole solution in Einstein Gauss Bonnet gravity in presence of a string cloud. We have found that there are three black holes in certain range of parameter for small enough chemical potential. As the chemical potential becomes greater than a critical value, it admits only a single black hole. The free enrgy can be computed from the on-shell Euclidean action, which requires a subtraction and in presence of string cloud it has an additional twist. We need to subtract the contribution of a  extremal black hole with zero charge. The study of free energy and specific heat reveals that the medium black hole is unstable.

It is interesting to study the  quak-antiquark potential in the dual gauge theory as appeared in  \cite{Yang, Li:2017tdz, Chakrabortty:2016xcb, Dey:2020yzl}. In the present case,  the quark-antiquark bound states  turns out to  exist only up to a maximum separation,  called screening length. Beyond the screening length, they get separated. The screening length is larger as the horizon radius of the black hole is smaller. So it is maximum for the small black hole, but it is finite and cannot be extended indefinitely. This indicates that the dual theory is in a deconfined phase. Nevertheless, here are new type of phase transition that occurs in higher derivative gravity theory \cite{Camanho:2012da} it would be interesting to explore that possibility in the present model.

The gauge/gravity duality has been applied in diverse areas that appear in condensed matter systems quite successfully. The present black hole solution is amenable to such applications. For example it can be used for study of different physical behaviours, such as a.c and d.c conductivities, thermoelectric properties, Fermi surfaces  etc. \cite{Donos:2014cya,Faulkner:2009wj}.
The present solution is electrically charged and one can try to obtain a similar solution, where it is magnetically charged as well and that can be useful for study of the magnetic properties. 

Another avenue to explore is to study the instabilities of the gravity solutions. In \cite{donos2} they studied possible instabilities around a broad class of magnetically charged solutions, with geometry AdS$_2$  which may give rise to spatially modulated states in the dual field theory and confirm finding the same. A similar study of instabilities of the present solution (or its magnetic analog) may lead to better understanding of the solution. 
\\

\noindent {\bf Note Added:} After submission of this work to arXiv, one paper \cite{Zhai:2023wig} appeared on arXiv which addresses similar issues as ours.

\section*{Acknowledgements}

TD thanks Rishi Pokhrel for fruitful discussions. 
SM thankfully acknowledges the assistance received from Science and Engineering Research Board (SERB), India (project file no. CRG/2019/002167).

\end{document}